\documentstyle[prb,aps,twocolumn]{revtex}
\begin{document}
\draft \flushbottom

\twocolumn[\hsize\textwidth\columnwidth\hsize\csname
@twocolumnfalse\endcsname

\title{Quantum-well states in ultrathin Ag(111) films deposited onto H-passivated Si(111)-(1x1) surfaces}
\author{A. Arranz,$^{1,2}$ J.F. S\'{a}nchez-Royo,$^{1,3}$ J. Avila,$^{1,4}$ V. P\'{e}rez-Dieste,$^{1,4}$
P. Dumas,$^{1}$ and M.C. Asensio$^{1,4,}$\cite{byline}}

\address{$^{1}$LURE, Centre Universitaire Paris-Sud, B\^{a}t. 209 D, B.P. 34, 91898
Orsay Cedex, France}
\address{$^{2}$Dpt. F\'{\i}sica Aplicada, Fac. de Ciencias, C-XII, Univ.
Aut\'{o}noma de Madrid, Cantoblanco, 28049 Madrid, Spain}
\address{$^{3}$Dpt. F\'{\i}sica Aplicada, ICMUV, Univ. de Valencia, c/Dr.
Moliner 50, 46100 Burjassot, Valencia, Spain}
\address{$^{4}$Instituto de Ciencia de Materiales de Madrid, CSIC,
Cantoblanco, 28049 Madrid, Spain}

\date{\today}
\maketitle
\begin{abstract}

Ag(111) films were deposited at room temperature onto H-passivated
Si(111)-(1x1) substrates, and subsequently annealed at 300
$^{o}$C. An abrupt non-reactive Ag/Si interface is formed, and
very uniform non-strained Ag(111) films of 6-12 monolayers have
been grown. Angle resolved photoemission spectroscopy has been
used to study the valence band electronic properties of these
films. Well-defined Ag \textit{sp} quantum-well states (QWS) have
been observed at discrete energies between 0.5-2eV below the Fermi
level, and their dispersions have been measured along the
$\overline{\Gamma}\overline{K}$, $\overline{\Gamma M (M')}$ and
${\Gamma}L$ symmetry directions. QWS show a parabolic
bidimensional dispersion, with in-plane effective mass of
0.38-0.50\textit{m$_{o}$}, along the
$\overline{\Gamma}\overline{K}$ and $\overline{\Gamma M (M')}$
directions, whereas no dispersion has been found along the
${\Gamma}L$ direction, indicating the low-dimensional electronic
character of these states. The binding energy dependence of the
QWS as a function of Ag film thickness has been analyzed in the
framework of the phase accumulation model. According to this
model, a reflectivity of 70\% has been estimated for the
Ag-\textit{sp} states at the Ag/H/Si(111)-(1x1) interface.

\end{abstract}

 ] %\narrowtext

\section{Introduction}

During the last years, the study of low-dimensional structures has
attracted considerable interest because spatial confinement of
electrons in thin films results in discrete quantum-well states
(QWS). In semiconductor layer systems, these effects are well
known, and they have been already used in electronic devices.
However, the observation of quantization effects in metallic
layers, has been restricted to a few systems. In this respect,
thin Ag films have attracted a great interest due to the nearly
free-electron characteristics of the \textit{sp} bands over large
regions of the Brillouin zone (BZ). Ag-\textit{sp} QWS have been
widely observed in thin Ag films deposited onto several metallic
substrates.\cite{ref1,ref2,ref3,ref4,ref5,ref6,ref7,ref8,ref9,ref10,ref11,ref12,ref13}
In contrast, confinement effects on Ag films deposited onto
semiconductor substrates have been only observed in a scarce
number of systems.\cite{ref14,ref15,ref16} Weak peaks associated
to QWS have been observed by Wachs \textit{et al}\cite{ref14} in
the photoemission spectra of 5-15 monolayers (ML) Ag films
deposited on Si(111)-(7x7) at room temperature (RT). The weakness
of the QWS observed by these authors could be associated with the
formation of a non-uniform Ag island size distribution, which
would wash out the manifestation of quantum size effects in
photoemission. These facts suggest that an improvement of Ag
deposition conditions and silicon substrate preparation could play
a key role for the observation of well-defined QWS in thin Ag
films deposited onto silicon substrates. In fact, this has been
observed by Neuhold and Horn,\cite{ref16} on a 21ML Ag film
deposited at 130 K onto a Si(111)-(7x7) substrate, and
subsequently annealed at RT. Such special deposition conditions
lead to the formation of more uniform Ag films, where quantum-size
effects have been observed by photoemission, although strain
induced by film growth was also present in the film.\cite{ref16}
Therefore, it would be desirable from both fundamental and
technological point of views, to consider alternative better
Si(111) surfaces free of complex reconstruction structures, which
could lead to the formation of high-quality non-strained Ag(111)
films.

The nucleation and growth mode of thin Ag films deposited onto Si
surfaces can be modified by changing the surface free energy of
the substrate. This modification can be accomplished by
termination of the Si(111) surface by a foreign
atom.\cite{ref17,ref18} By this reason, there is a great interest
in the formation of high quality artificially H-terminated
Si(111)-(1x1) surfaces, by both a wet chemical treatment and an
atomic hydrogen-based method.\cite{ref19} Hydrogenation of the
Si(111)-(7x7) surface saturates the dangling bonds, restoring the
(1x1) symmetry of bulk silicon. In a recent study on the
electronic structure of the H-terminated Si(111)-(1x1) surface,
the high quality of such a type of unreconstructed Si surfaces
prepared by a wet chemical treatment has been confirmed by angle
resolved photoemission spectroscopy (ARPES).\cite{ref20} Four
remarkably sharp features have been resolved in the valence band
photoemission spectra at the $\overline{K}$-point, being
attributed to: (i) a surface resonance with a p$_{x}$-p$_{y}$
symmetry, (ii) a surface state identified as a Si-Si backbond
state, and (iii) two higher binding-energy H-Si surface states.

The growth mode and structure of thin Ag films deposited onto
hydrogen passivated silicon surfaces are substantially different
from the observed ones for Si(111)-(7x7)
surfaces.\cite{ref17,ref18} When Ag is deposited onto a clean
Si(111)-(7x7) surface, the growth proceeds in a
quasi-layer-by-layer mode  at RT, and according to
Stranski-Krastanov growth mode at 300 $^{o}$C. Opposite to this,
Ag deposition at 300 $^{o}$C onto the H/Si(111)-(1x1) surface,
leads to a quasi-layer-by-layer growth mode. In this case, the
formation of islands much thinner and with a narrower size
distribution than those on the non-passivated substrate has been
observed by impact-collision ion-scattering spectroscopy
(ICISS)\cite{ref17} and scanning electron microscopy
(SEM)\cite{ref18,ref21} . In addition, Sumitomo \textit{et
al}\cite{ref17}, have found that Ag films deposited onto
Si(111)-(7x7) substrates at RT have a two-domain Ag(111) islands
distribution, whereas those grown on the hydrogen-terminated
surface at 300 $^{o}$C have a single-domain preferred orientation.
According to these authors, hydrogen-mediated epitaxy (HME) of
single-domain Ag(111) films deposited on Si(111)-(7x7) substrates
at 300 $^{o}$C is observed. It has been suggested that hydrogen at
the interface can be partially removed during Ag deposition at
high temperatures.\cite{ref22,ref23,ref24}  If the hydrogen
coverage at the interface would decrease below 0.3 ML, an
undesired reconstruction of the surface, to a 2D layer of the
Si(111)$\sqrt{3}$x$\sqrt{3}$-Ag structure plus thicker Ag(111)
islands could occur.\cite{ref22,ref25} So, it would be desirable
to lower the Ag temperature deposition in order to avoid a
significant hydrogen elimination at the interface, and therefore
the above-mentioned surface reconstruction.

Ag deposition at RT onto hydrogen-terminated Si(111) surfaces has
been only studied by scanning tunneling microscopy
(STM).\cite{ref25} It was observed an increase of the average size
of the Ag islands upon increasing the substrate temperature from
RT to 300 $^{o}$C. On the contrary, the Ag(111) island density
decreases, and therefore RT deposition leads to even more uniform
and compact Ag films than those deposited at higher temperatures.
Unfortunately, in the literature there is a lack of information
from structural techniques about the single or two-domain
preferred orientation of thin Ag films deposited on
hydrogen-terminated Si substrates at RT. Recently, the
above-mentioned subject has been addressed studying the Fermi
surface (FS) of 6ML Ag films deposited onto H-passivated
Si(111)-(1x1) surfaces at RT, and subsequently annealed at 300
$^{o}$C.\cite{ref26} This study has shown that the measured FS
reflects a sixfold symmetry rather than the threefold symmetry
expected for a Ag(111) single crystal. This behavior has confirmed
the fact that these Ag films are composed by two domains rotated
60$^{o}$, being the first evidence of a two-domain preferred
orientation of the Ag(111) films deposited at RT onto H-terminated
Si(111)-(1x1) surfaces.

In this work, thin Ag(111) films have been deposited in
ultra-high-vacuum (UHV) at room temperature (RT) onto H-passivated
Si(111)-(1x1) substrates, and subsequently annealed at 300 $^{o}$C
to enhance the film uniformity. In such a way, an abrupt
non-reactive Ag/Si interface is formed, and high-quality
non-strained thin Ag(111) films of 6-12ML have been obtained. The
quality of the silver films has been probed by ARPES. Well-defined
Ag \textit{sp} QWS have been observed at discrete energies between
0.5-2eV below the Fermi level ($E_{F}$), and their energy
dispersions have been measured along the
$\overline{\Gamma}\overline{K}$, $\overline{\Gamma M (M')}$ and
${\Gamma}L$  symmetry directions. QWS show a parabolic dispersion,
with in-plane effective mass of 0.38-0.50\textit{m$_{o}$}, along
the $\overline{\Gamma}\overline{K}$ and $\overline{\Gamma M (M')}$
symmetry directions, whereas no dispersion has been found along
the ${\Gamma}L$ direction, supporting the two-dimensional
electronic character of these states. On the other hand, the
binding energy dependence of the QWS as a function of Ag film
thickness has been analyzed in the framework of the phase
accumulation model.\cite{ref3} A good agreement between
experimental data and the above-mentioned model is obtained for
the Ag/H/Si(111)-(1x1) system.

\section{Experimental details}

The experiments were performed at LURE (Orsay, France) using the
French-Spanish (PES2) experimental station of the Super-Aco
storage ring, described elsewhere.\cite{ref27} The measurements
were carried out in a purpose-built UHV system, with a base
pressure of 5x10$^{-11}$ mbar, equipped with an angle resolving 50
mm hemispherical VSW analyzer coupled on a goniometer inside the
chamber. The manipulator was mounted in a two-axes goniometer that
allows rotation of the sample in: (i) The whole 360$ ^{o}$
azimuthal angle ($\phi$) and (ii) In the 180$^{o}$ polar emission
angle relative to surface normal ($\theta$), with an overall
angular resolution of 0.5$^{o}$. The current incident angle of the
light ($\Theta_{i}$) was 45$^{o}$.The available energy of light
($h\nu$) was between 18 and 150 eV.

The substrates were n-doped Si(111) single crystal, with a nominal
resistivity of 100 $\Omega$cm. It was prepared \textsl{ex situ}
using a wet chemical treatment that results in a passivated
H/Si(111)-(1x1) surface.\cite{ref19} After introducing the
substrate in the analysis chamber, the quality of the surface was
checked out through the sharpness of the features appearing in the
valence band photoemission spectrum at $\overline{K}$-point, which
are attributed to intrinsic H-Si surface states.\cite{ref20} Ag
was evaporated onto the surface at RT. The rate of evaporation,
0.06 ML/min, was determined by using a quartz microbalance. In
these conditions, Ag films of thickness ranging from 6 to 12 ML
were deposited onto H/Si(111)-(1x1) surfaces at RT, and
subsequently annealed at 300 $^{o}$C to enhance the film
uniformity.

\section{Results and Discussion}

Normal and off-normal valence band (VB) photoemission spectra,
after background subtraction, of 6-12ML Ag films deposited at RT
onto H-terminated Si(111)-(1x1) substrates and subsequently
annealed at 300 $^o$C, are presented in order to obtain the Ag
\textit{sp}-band dispersion along the ${\Gamma}L$,
$\overline{\Gamma}\overline{K}$ and $\overline{\Gamma M (M')}$
symmetry directions. Fig. 1 shows the evolution of the normal
emission VB spectra measured with $h\nu$=32eV as a function of Ag
thickness. The intense and sharp peak observed at 0.1eV below
E$_F$ is the surface state (SS) in the band gap along the
direction of the Ag(111) crystal. The sharpness of this peak
reflects the good crystallinity and orientation of the Ag film. It
should be pointed out that the sharpness observed for this peak is
comparable to the observed one for Ag films deposited onto highly
oriented pyrolytic graphite, HOPG(0001), where a very narrow
height distribution of Ag islands has been
observed.\cite{ref4,ref16} Neuhold and Horn,\cite{ref16} have
attributed the depopulation of the Ag SS to the presence of
tensile strain induced by film growth in a 21ML Ag film on
Si(111)-7x7. Comparison of our results with the works of Wachs
\textit{et al}\cite{ref14} and Neuhold and Horn,\cite{ref16}
suggests an improvement of the quality of the silver films, as a
consequence of the ideally H-terminated Si(111) substrate surface
used in this work. As a consequence, high-quality non-strained Ag
films can be successfully prepared onto H-passivated Si(111)-(1x1)
surfaces at RT, with a subsequent annealing. Alternative
techniques for Ag deposition based on the cooling of the substrate
to 130 K and subsequent annealing at RT, lead also to the
formation of high-quality Ag films where manifestation of
quantum-size effects in photoemission has been
observed.\cite{ref16} However, effects associated to the strain
cannot be avoided, and it should be taken into account to describe
the electronic properties of such  Ag thin films.

Three series of well-resolved intense Ag \textit{sp} QWS,
characterized by the quantum number $\nu$=1 to 3, can be observed
at discrete energies between 0.5-2eV below the Fermi level in Fig.
1. The evolution of the QWS energies as a function of Ag thickness
for the different series observed ($\nu$=1, 2 and 3) will be
analyzed later in the backdrop of the phase accumulation model,
(PAM).\cite{ref3} The quantum number $\nu$ is defined as,
$\nu=m-n$, where $m$ is the number of Ag MLs, and $n$ is the
number of antinodes in the probability density. Ag \textit{sp} QWS
of similar quality have also been observed by Neuhold and
Horn\cite{ref16} in a 21ML Ag film on Si(111)-7x7, deposited at
130 K and subsequently annealed at RT, suggesting that special Ag
deposition conditions, or as shown in this work, Si(111)
substrates free of complex surface reconstructions, are necessary
in order to observe quantum size effects in Ag films deposited
over Si(111) substrates. Therefore, it should be pointed out that
the weakness of the QWS peaks observed by Wachs \textit{et
al}\cite{ref14} in the photoemission spectra of 5-15 MLs Ag films
deposited at RT onto Si(111)-(7x7) should be associated with the
formation of a non-uniform Ag island size distribution, as a
consequence of standard deposition at RT. In the lower part of
Fig. 1, the \textit{sp}-band dispersion, E($k_\bot$), of bulk Ag
along the ${\Gamma}L$ direction is shown in order to illustrate
the physical origin of the QWS peaks observed in the spectra. The
above-mentioned dispersion relation E($k_\bot$) has been simulated
by the two band model given by Eq. (1),\cite{ref4}

\begin{eqnarray}\label{equation1}
  {E(k_\bot)=E_o-A(k_{BZ}-k_\bot)^2} \nonumber\\
   {+ \,\, U - \sqrt{4A^2B(k_{BZ}-k_\bot)^2 + U^2}}
\end{eqnarray}
where $A=\hbar^2/2m^*$, $B=3\pi^2/a^2$, $a$=4.09\AA$ $ is the
silver lattice constant, $U$=4.2eV is the width of the gap at the
$L$ symmetry point of Ag(111), $E_o$=0.31eV is the position of the
\textit{sp}-band edge relative to E$_F$, $m^*$=0.7\textit{m$_{o}$}
is the effective mass of the electrons in this band, being
\textit{m$_{o}$} the free electron mass, and
$k_{BZ}$=1.33\AA$^{-1}$ is the wave vector at the BZ boundary (the
$L$ point in the [111] direction).

Angle-resolved photoemission spectra of the VB measured with
$h\nu$=32eV, along the $\overline{\Gamma}\overline{K}$ and
$\overline{\Gamma M (M')}$ symmetry directions are presented in
Figs. 2 and 3, for 6 and 7ML Ag films, respectively. As a
reference, a scheme of the Ag(111) surface reciprocal-lattice unit
cell is shown in the figures. It should be pointed out, that
$\overline{M}$ and $\overline{M'}$ points of the surface BZ
overlap because Ag films are composed by two 60$^o$ rotated
domains.\cite{ref26} Dashed lines have been added to show the
energy dispersion of the different peaks observed in the spectra.
These peaks are the QWS denoted by $\nu$=1 and $\nu$=2 in Fig. 1,
and the feature associated with the Ag(111) surface state. The
insets of figs. 2 and 3 show the band diagram extracted from the
dispersion of these features. Solid lines are parabolic fits to
Eq. (2), that are expected to be a good approximation for small
parallel wave vector, $k_\|$, values.\cite{ref2}

\begin{equation}\label{equation2}
  E_\nu(k_\|)=E_\nu(k_\|=0) + \hbar^2k_\|^2/2m_\|^*
\end{equation}
In Eq. (2), $E_\nu(k_\|=0)$ and the in-plane effective mass,
$m_\|^*$, are fitting parameters.

In agreement with a previous work,\cite{ref26} two \textit{sp} Ag
derived surface states, \textit{SS1} and \textit{SS2}, at binding
energies of $\sim$ 0.1 and 0.32eV, respectively, are found in the
gap along the ${\Gamma}L$ direction of the Ag(111) films. The
existence of two different surface states with the same origin was
attributed to the inhomogeneous presence of hydrogen at the
interface, due to the fact that annealing process partially
removes hydrogen, and therefore, a downshift of the Ag surface
state is produced in regions where hydrogen still remains at the
interface. It should be noted, that for Ag film thicknesses $\geq$
6ML, the \textit{SS2} feature appears as an asymmetry of the more
intense \textit{SS1} peak in normal emission spectra. However,
\textit{SS2} peak becomes clearly resolved in off-normal emission
spectra of Figs. 2 and 3, as a consequence of the Fermi level
crossing of the \textit{SS1} state.

Surface states and QWS show a parabolic in-plane dispersion as can
be observed in the insets of figs. 2 and 3. The QWS in-plane
effective mass, $m_\|^*$, has been obtained from the fit of
experimental data to Eq. (2), giving values between
0.38-0.50\textit{m$_{o}$}. These values are in good agreement with
the reported ones in the literature for silver films on Cu(111)
substrates, and are almost the same as the effective mass for the
Ag \textit{sp} valence band derived from a band structure
calculation.\cite{ref2,ref8} Moreover, it has been also observed
the linear increase of the in-plane effective mass with the
binding energy found by Mueller \textit{et al} for the Ag/Cu(111)
quantum-well system.\cite{ref2}

The evolution of the normal emission BV spectra as a function of
$h\nu$ for a 7ML Ag(111) film is presented in Fig. 4. As can be
observed, a negligible dispersion is found for the QWS along the
${\Gamma}L$ symmetry direction, as a consequence of the
quantization of the perpendicular wave vector component, $k_\bot$,
due to the finite and homogeneous thickness of the film. The
non-dispersive behavior along the ${\Gamma}L$ direction, and the
parabolic in-plane dispersion along the
$\overline{\Gamma}\overline{K}$ and $\overline{\Gamma M (M')}$
symmetry directions of the QWS and surface states, indicate the
two-dimensional electronic character of these states.

Observation of confinement effects in the
Ag/H/Si(111)-(1x1)system, requires either the presence of a
relative gap in the Si(111) substrate, or the presence of a
so-called symmetry or hybridization gap, in order to avoid the
coupling of Ag-\textit{sp} states with Si(111) \textit{sp}-states
of similar symmetry.\cite{ref1} Well-defined QWS have been
observed in this work for energies between 0.5 and 2eV below
$E_F$. Neuhold and Horn,\cite{ref16} have also observed similar
quality QWS up to an energy of 3eV below $E_F$. So, the
above-mentioned requirements should be fulfilled for hydrogen
passivated and non-passivated silicon substrates. It should be
pointed out, that according to the calculated valence band
structure,\cite{ref20} the gap at the $\overline{\Gamma}$-point in
the [111] direction of the silicon substrate extends up to $\sim$
1eV below $E_F$, so the observation of QWS for binding energies
above 1eV, should be explained by the lack of coupling between the
Ag-\textit{sp} states and the Si(111) bulk states. The role of the
presence of hydrogen at the Ag/Si(111) interface should be also
taken into account. It could act as a barrier, that would enhance
the inhibition of the above-mentioned coupling, as it has been
observed in the 15ML Ag + $x$ML Au + Ag(111) quantum-well system,
upon increasing the Au interlayer thickness from  $x$=0 to
3ML.\cite{ref6}

In order to further analyze the dependence of the QWS energies
with the Ag film thickness, the binding energies of the QWS
observed in Fig. 1 are plotted (solid symbols) as a function of
the Ag film thickness in Fig. 5. In this figure, data
corresponding to 5-13ML Ag films deposited onto Si(111)-7x7 at
RT\cite{ref14} have been also plotted for comparison (open
symbols). According to the phase accumulation model,\cite{ref3}
the energy position of the QWS is given by Eq. (3),

\begin{equation}\label{equation3}
  2k_{\bot}(E)d + \Phi_C(E) + \Phi_B(E) = 2\nu\pi
\end{equation}
where $k_\bot(E)$ is the electron wave-vector component normal to
the surface, $d$ is the Ag film thickness, $\nu$  is the quantum
number previously defined, and $\Phi_C(E)$ and $\Phi_B(E)$  are
the phase shifts upon reflection of the electron wave-function at
the film/substrate and film/vacuum interfaces, respectively. The
above-mentioned model considers that QWS are electron waves
trapped inside the Ag film between the barriers at the Ag/Si(111)
and Ag/vacuum interfaces. Upon reflection of the Ag-\textit{sp}
electrons at the interfaces, phase changes are introduced,
$\Phi_C(E)$ and $\Phi_B(E)$, whereas $k_{\bot}(E)d$ is the phase
change accumulated upon traversing the Ag film. In this way, thin
uniform Ag films can be considered as an electron
interferometer,\cite{ref29} where the electron undergoes multiple
reflections between the interfaces. For a given Ag film thickness,
a Ag-\textit{sp} electron standing wave is expected at a discrete
energy in the photoemission spectra, when the total phase change
satisfies the condition given by Eq. (3). Fig. 6 shows a schematic
diagram illustrating the analogy of the phase accumulation model
with an electron interferometer for the Ag/Si(111)-(1x1)-H system.
Multiple reflection at the interfaces of Ag-\textit{sp} electrons
are indicated by dashed arrows. Ideal confinement is achieved for
a 100\% interface reflectivity. A reflectivity lower than 100\%
would imply partial transmission of the electron wave outside the
Ag film as evanescent states, indicated by short arrows in the
figure.

In Eq. (3), the dispersion relation $k_\bot(E)$ for Ag(111) in the
${\Gamma}L$ direction can be simulated by the two-band model
previously defined by Eq. (1). $\Phi_B(E)$ and $\Phi_C(E)$, are
described by Eqs. (4) and (5), respectively,

\begin{equation}\label{equation4}
  \Phi_B(E) =\pi\Biggl[{\sqrt{\frac{3.4eV}{E{_V}-E}}-1}\Biggr]
\end{equation}

\begin{equation}\label{equation5}
  \Phi_C(E)= - \alpha\pi
\end{equation}
where $E_V$=3.7eV is the vacuum level,\cite{ref28} and $\alpha$ is
a fitting parameter that can take values between 0 and 1. Eq. (4)
is the common empirical approximation used in the literature for
$\Phi_B(E)$,\cite{ref3} whereas $\Phi_C(E)$ has been defined to
take values between 0 and $-\pi$. A value of $\alpha$=1 in Eq.
(5), would imply a nearly 100\% reflectivity at the Ag/substrate
interface, as has been proposed for the Ag/fcc-Fe(100)
system.\cite{ref13} For other systems,\cite{ref3,ref4} a different
empirical relation based on a step potential approximation for
$\Phi_C(E)$ has been proposed, but this does not seem to be valid
for the Ag/Si system, where a combination of relative and symmetry
gaps should be considered in order to account for the observed
confinement effects.

Dashed lines in Fig. 5 represent the calculated thickness
dependence of the binding energy of the QWS (with $\nu$=1-3) based
on the fit of the experimental data (solid symbols) to the PAM
proposed by Eq. (3). The best fit has been obtained for
$\alpha$=0.7. This value can be interpreted as a $\sim$ 70\%
reflectivity of the Ag-\textit{sp} states at the Ag/H/Si(111)
interface. A good agreement between experimental data and
calculated lines is obtained, suggesting that the PAM can
successfully characterize the QWS observed in the
Ag/H/Si(111)-(1x1) system, taken into account that an independent
energy approximation for $\Phi_C(E)$ has been used. Obviously, a
better agreement could have been obtained if the total phase shift
would have been simulated by a third-order
polynomial,\cite{ref11,ref12} but in this case direct information
about the Ag-\textit{sp} states reflectivity at the interface
would have been lost.

Finally, it should be pointed out, that the same model of Eq. (3)
applied to data of the Ag/Si(111)-(7x7) system\cite{ref14} (open
symbols of Fig. 5) gives $\alpha\sim$ 0. This would imply a nearly
zero reflectivity at the Ag/Si interface, and therefore no QWS
should have been observed. In order to account for this result,
two additional effects should be taken into account. The influence
of induced-film-growth strain in the electronic structure ($E(k)$)
of Ag films deposited onto Si(111)-7x7 substrates at
RT,\cite{ref16} and the inhibiting electronic-coupling role of the
remaining hydrogen at the interface.

\section{Summary and Conclusions}

Thin Ag(111) films have been deposited in UHV at room temperature
onto H-passivated Si(111)-(1x1) substrates, and subsequently
annealed at 300 $^{o}$C to enhance the film uniformity. Deposition
onto H-terminated Si(111)-(1x1) unreconstructed surfaces supresses
strain induced by film growth. As a consequence, an abrupt
non-reactive Ag/Si interface is formed, and very uniform
non-strained Ag(111) films of 6-12ML have been obtained. The
quality of the silver films has been probed by ARPES. Well-defined
Ag-\textit{sp} QWS have been observed at discrete energies between
0.5-2eV below the Fermi level, and their energy dispersion have
been measured along the $\overline{\Gamma}\overline{K}$,
$\overline{\Gamma M (M')}$ and ${\Gamma}L$  symmetry directions.
QWS show a parabolic dispersion, with in-plane effective mass of
0.38-0.50\textit{m$_{o}$}, along the
$\overline{\Gamma}\overline{K}$ and $\overline{\Gamma M (M')}$
directions, whereas no dispersion has been found along the
${\Gamma}L$ direction, supporting the two-dimensional electronic
character of these states.

On the other hand, the binding energy dependence of the QWS as a
function of Ag film thickness has been analyzed in the framework
of the phase accumulation model. A good agreement between
experimental data and the above-mentioned model is obtained for
the Ag/H/Si(111)-(1x1) system, and a $\sim$70\% reflectivity for
the Ag-\textit{sp} states at the Ag film/substrate interface has
been derived. Comparison of our results with the scarce works
existing in the literature on QWS in the Ag/Si(111) system,
suggests that the influence of strain induced by film growth in
the electronic structure of Ag films deposited onto Si(111)-7x7
substrates, should be also taken into account in order to
accurately describe the electronic properties of those films. In
addition, hydrogen at the interface enhances the degree of
confinement of electrons in the Ag thin film, acting as a barrier
that inhibits the electronic-coupling between Ag and Si
\textit{sp} states at the interface.

\acknowledgments

This work was financed by DGICYT (Spain) (Grant No. PB-97-1199) and the
Large Scale Facilities program of the EU to LURE. Financial support from the
Comunidad Aut\'{o}noma de Madrid (Project No. 07N/0042/98) is also
acknowledged. A.A. and J.F.S.-R. acknowledge financial support from the
Ministerio de Educaci\'{o}n y Cultura of Spain.

%%%%%%%%%% FIGURES

\begin{figure}
\caption{Normal emission valence band spectra measured with
$h\nu$=32eV for different Ag film thicknesses deposited onto an
H-terminated Si(111)-(1x1) substrate at RT, and subsequently
annealed at 300 $^{o}$C. In the bottom, the \textit{sp}-band
dispersion, E($k_\bot$), of bulk Ag along the ${\Gamma}L$
direction is shown in order to illustrate the physical origin of
the QWS peaks observed in the spectra.} \label{fig1}
\end{figure}

\begin{figure}
\caption{Angle-resolved photoemission spectra of the VB measured
with $h\nu$=32eV along the $\overline{\Gamma}\overline{K}$
symmetry direction for the 6ML Ag film of Fig. 1.  As a reference,
a scheme of the Ag surface reciprocal-lattice unit cell is given
in the figure. The inset shows the band diagram extracted from the
dispersion of the features observed in the spectra.} \label{fig2}
\end{figure}

\begin{figure}
\caption{Angle-resolved photoemission spectra of the VB measured
with $h\nu$=32eV along the $\overline{\Gamma M (M')}$ symmetry
direction for the 7ML Ag film of Fig. 1. As a reference, a scheme
of the Ag surface reciprocal-lattice unit cell is given in the
figure. The inset shows the band diagram extracted from the
dispersion of the features observed in the spectra.} \label{fig3}
\end{figure}

\begin{figure}
\caption{Normal emission valence band spectra as a function of the
energy of light, $h\nu$, for the 7ML Ag film of Fig. 1.}
\label{fig4}
\end{figure}

\begin{figure}
\caption{QWS binding energies of Fig. 1 as a function of Ag film
thickness (solid symbols). For comparison data corresponding to
5-13ML Ag films deposited onto Si(111)-(7x7) at RT, extracted from
reference 14, have been also plotted (open symbols). Dashed lines
are the calculated thickness dependence of the QWS binding
energies according to the phase accumulation model proposed in the
text.} \label{fig5}
\end{figure}

\begin{figure}
\caption{Schematic diagram illustrating the analogy of the
multiple-reflection phase accumulation model with an electron
interferometer for the Ag/Si(111)-(1x1)-H system. Multiple
reflections at the interfaces of Ag-\textit{sp} electrons are
indicated by dashed arrows. Evanescent states outside the Ag films
due to interface reflectivities lower than 100\%, are indicated by
short arrows.} \label{fig6}
\end{figure}

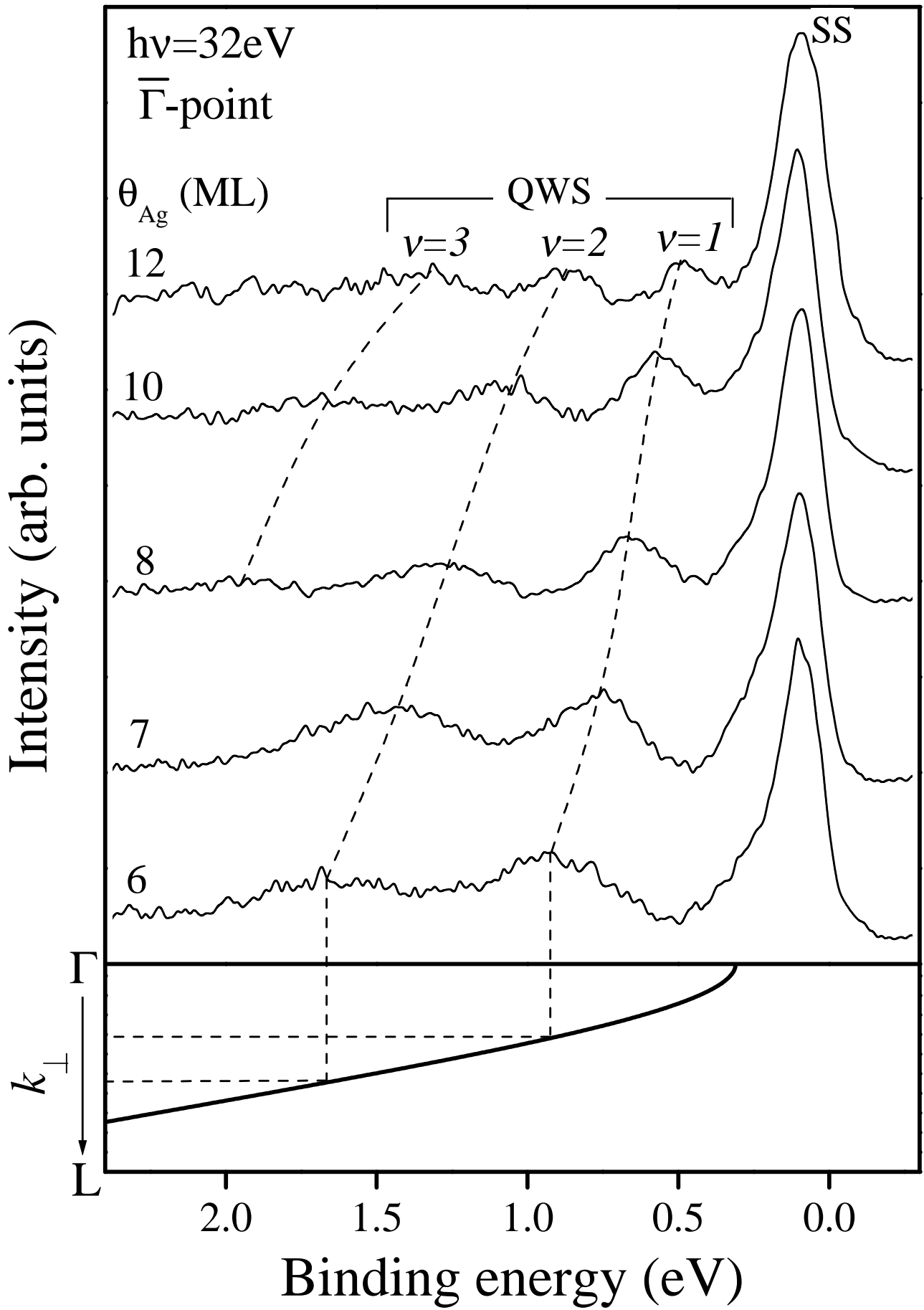
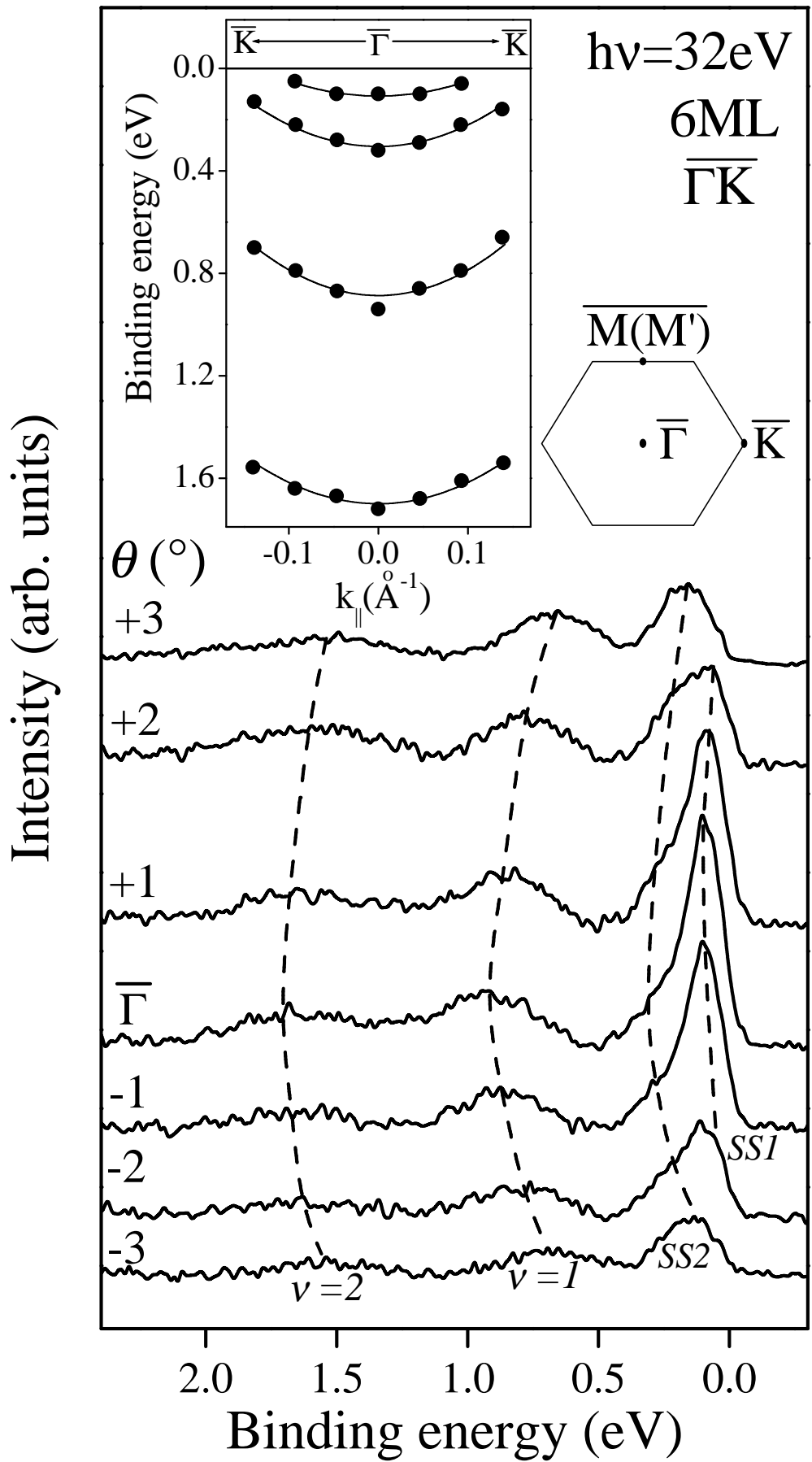
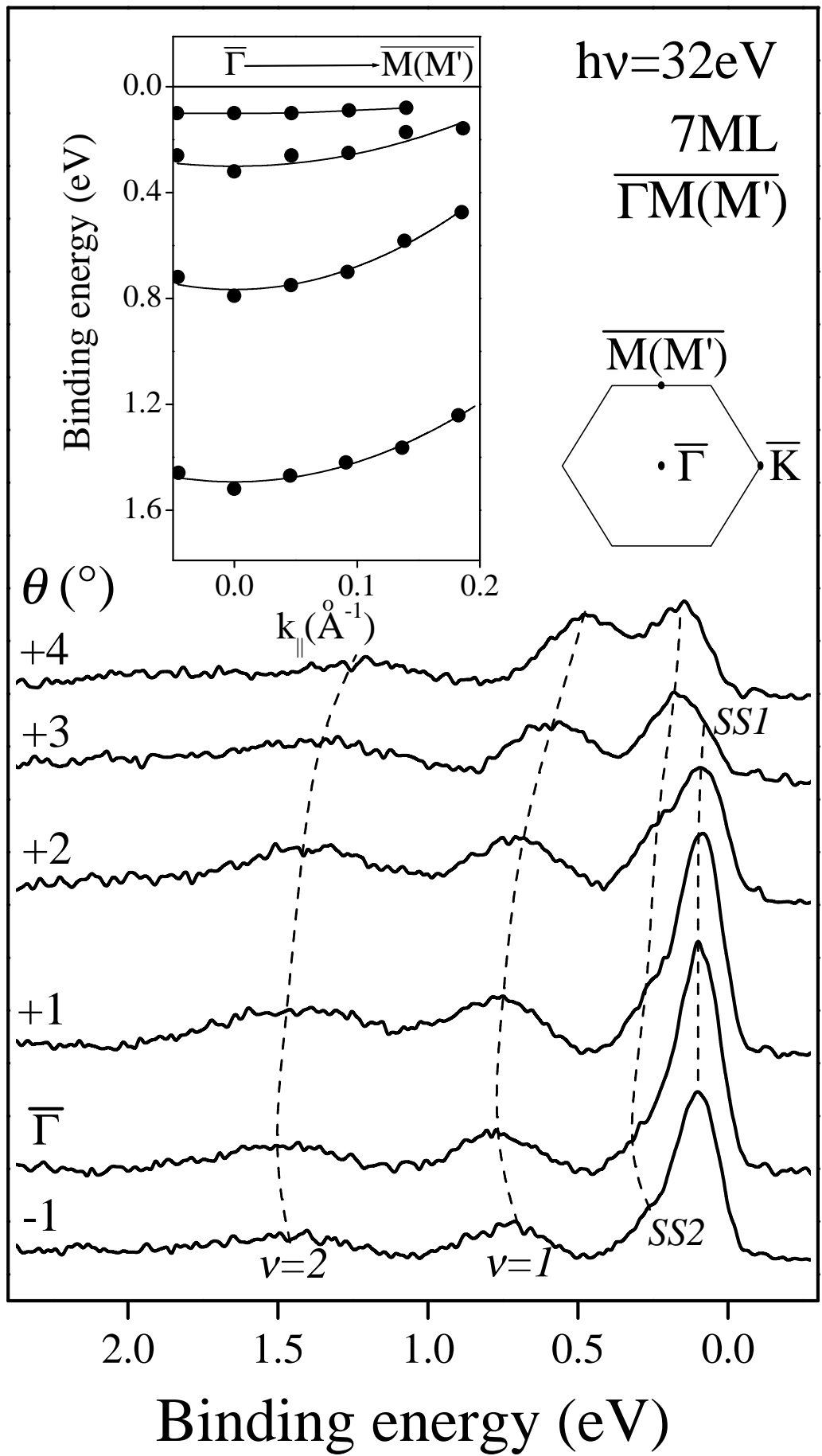
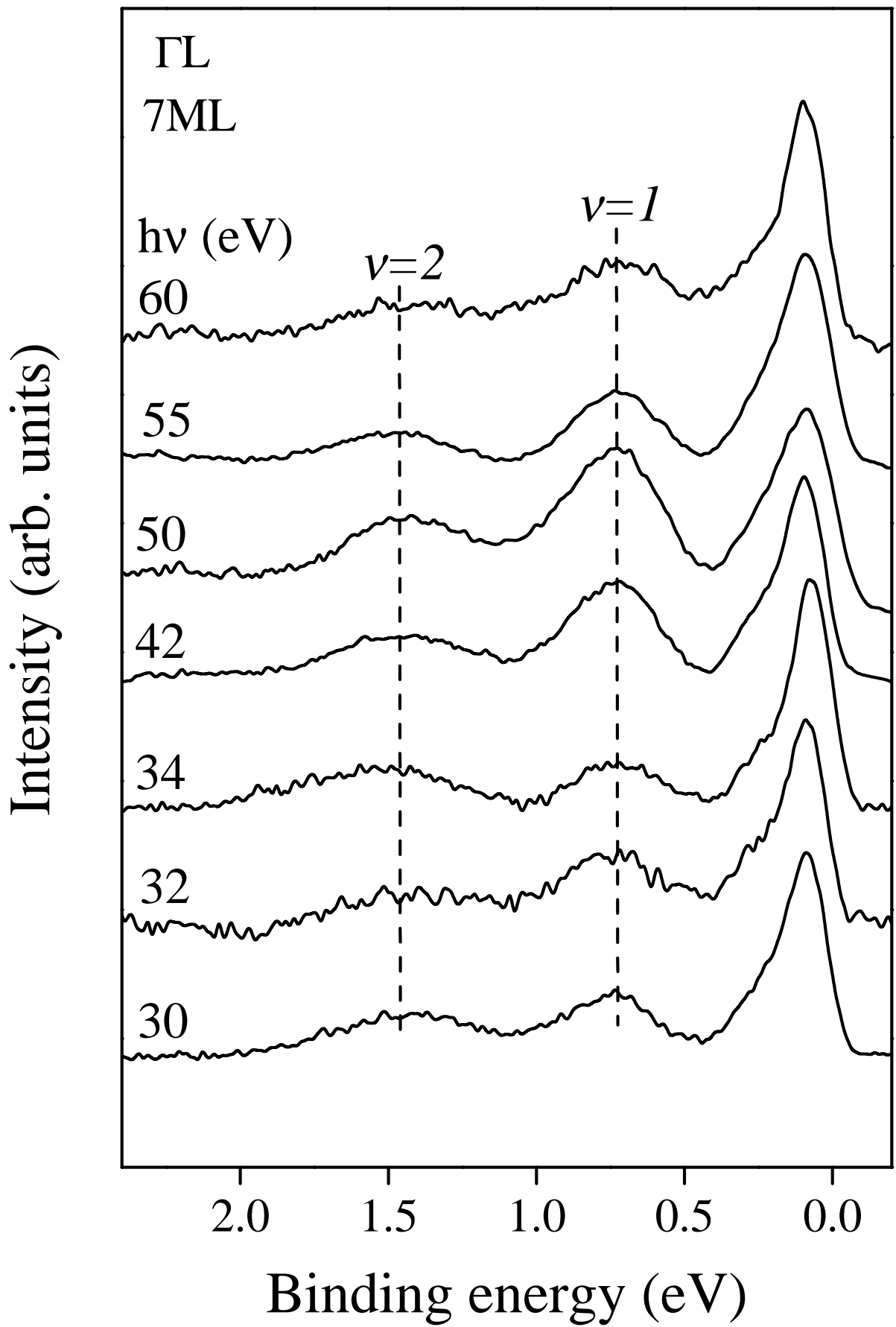
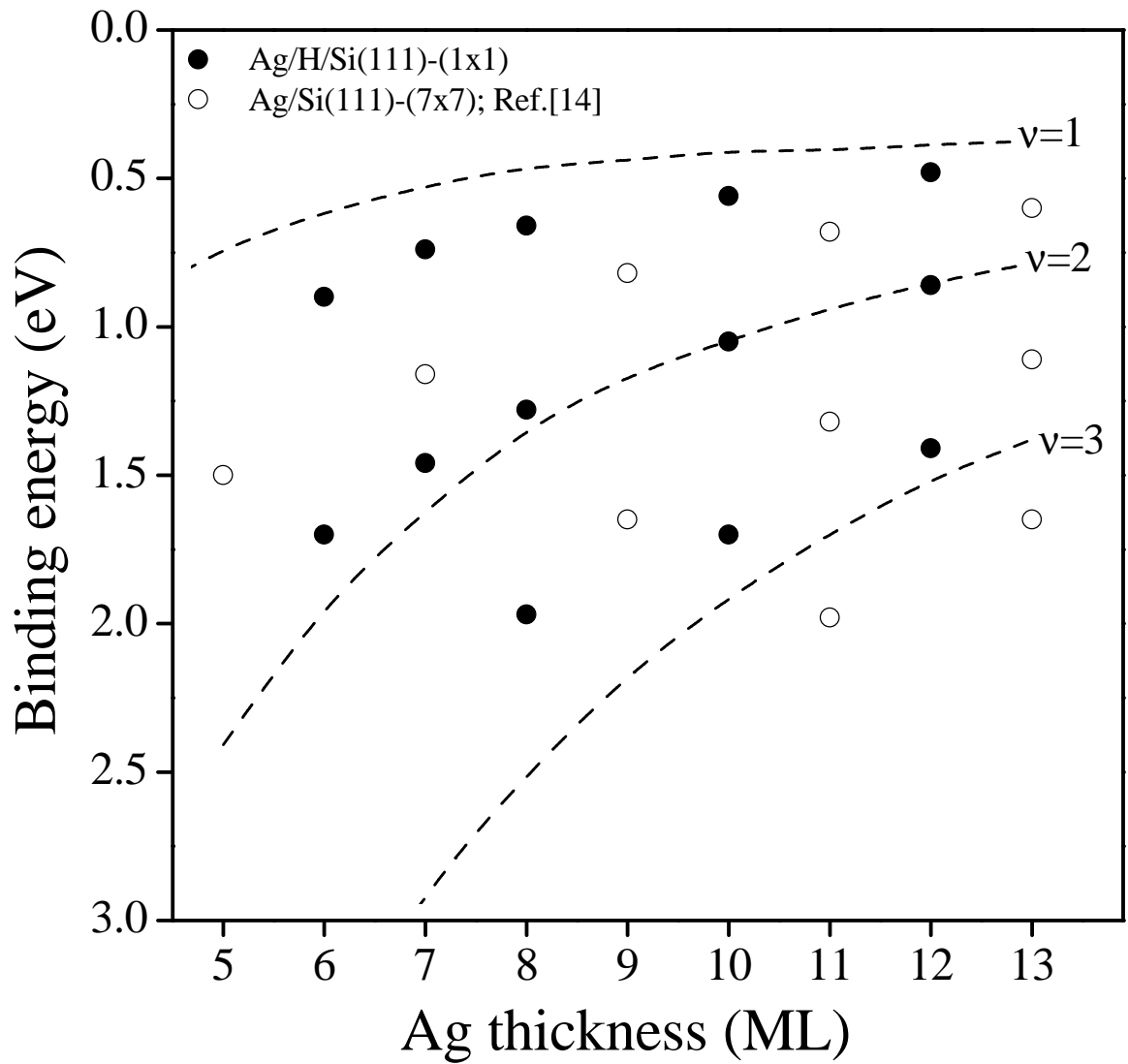
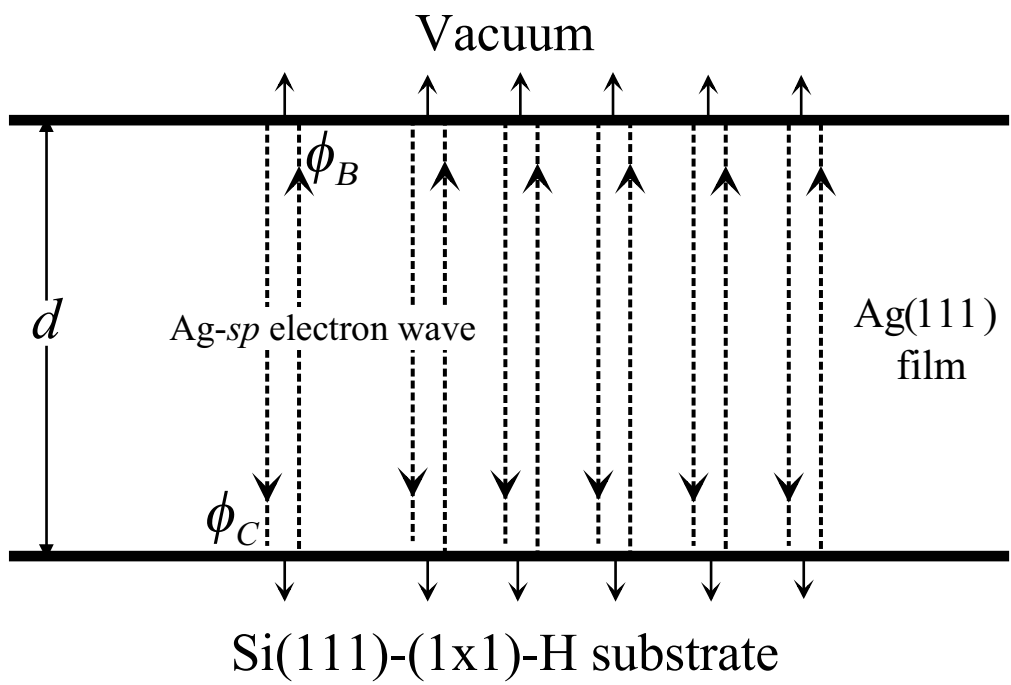
\end{document}